\documentclass[aps,showpacs,amsmath,amssymb,superscriptaddress,pra,twocolumn,longbibliography]{revtex4-1}

\usepackage{graphicx}
\usepackage{dcolumn}
\usepackage{bm}
\usepackage[latin1]{inputenc}
\usepackage{epstopdf}
\usepackage{amsmath}
\usepackage{amssymb}
\usepackage{latexsym}
\usepackage{mathrsfs}

\begin{document}

 
\title{Stochastic heating and self-induced cooling in optically bound pairs of atoms}

\author{Angel T. Gisbert} 
\affiliation{Dipartimento di Fisica "Aldo Pontremoli", Universit\`a degli Studi di Milano, Via Celoria 16, Milano I-20133, Italy}
\email{angel.tarramera@unimi.it}
\author{Nicola Piovella} 
\affiliation{Dipartimento di Fisica "Aldo Pontremoli", Universit\`a degli Studi di Milano, Via Celoria 16, Milano I-20133, Italy}
\email{Nicola.Piovella@unimi.it}
\author{Romain Bachelard} \affiliation{Universidade Federal de S\~{a}o
Carlos, Rod. Washington Luis, km 235, S/n - Jardim Guanabara, S\~{a}o
Carlos - SP, 13565-905, Brazil}
\email{bachelard.romain@gmail.com}

\date{\today}

\begin{abstract}
The light scattered by cold atoms induces mutual optical forces between them, which can lead to bound states. In addition to the trapping potential, this light-induced interaction generates a velocity-dependent force which damps or amplifies the stretching vibrational mode of the two-atom ``molecule''. 
This velocity-dependent force acts on time scales much longer than the mode period or the dipole dynamics, determining the true stability of the bound state.
We show that for two atoms, the stochastic heating due to spontaneous emission always exceeds the bounding effect, so pairs of cold atoms cannot be truly stable without an extra cooling mechanism.
\end{abstract}

\maketitle

\section{Introduction}

The advent of the laser and the subsequent cooling techniques applied to atomic samples have been a fundamental tool to lower their temperature by many orders of magnitudes~\cite{Phillips1998}. Eventually, temperatures can be reached where the Doppler effect has a negligible role, and coherences between the atoms can be preserved over the size of the sample. The Bose-Einstein condensation was a major step in this direction~\cite{Anderson1995}, which gave access to several new phases of matter, both for disordered systems and ordered systems (such as the Mott insulating phase when ultracold atoms are trapped into optical lattices~\cite{Greiner2002}). Apart from sympathetic cooling~\cite{Myatt1997}, cooling techniques do not involve interactions between the atoms, but rather between the laser photons and independent atoms. The atoms are thus cooled independently, and the atomic sample is spatially confined by a quasi-harmonic potential. 

Yet light-induced interactions between the atoms can be a powerful tool to create ordered systems~\cite{Labeyrie2014}. A paradigmatic example of cooperation in cold atoms is the collective atomic recoil lasing \cite{Bonifacio1994,Slama2007} observed when a cold or ultracold atomic gas in an optical ring cavity is illuminated by an intense far-off-resonance laser beam, causing a self-induced density grating in the atomic sample. More generally, the optical dipole force on the atoms in a high-finesse optical cavity, together with the back-action of atomic motion onto the light field, gives rise to nonliner collective dynamics and self-organization \cite{Ritsch2013}.
All these schemes with atoms in optical resonators rely on the creation of an optical lattices generated by the atoms.

In a similar fashion, it has recently been proposed to optically bind pairs of atoms confined in two dimensions by a stationary wave, where each atom remains at a multiple of the optical wavelength from the other~\cite{Maximo2018}. This effect stems from the generation of a non--trivial potential landscape due to the interference between the trapping beams and the wave radiated by each atom (see Fig.\ref{fig:scheme}). As for atoms trapped in a one-dimensional optical lattice, the distance between the atoms is a multiple of the optical wavelength, as is well-known from optical binding with dielectrics~\cite{Burns1989,Burns1990}.

Nevertheless, differently from the optical binding of dielectrics which are immersed in a fluid to confine them~\cite{Grzegorczyk2006,Metzger2006,Karsek2006,Metzger2007,Dholakia2010,Grzegorczyk2014}, cold atoms are manipulated at ultralow pressure, so the surrounding medium can be considered to be vacuum. An important consequence pointed out in Ref.~\cite{Maximo2018} is that since each atom exerts a central force on the other, the angular momentum is preserved, instead of being damped by viscous forces as for dielectrics in fluids~\cite{Ng2005}. Yet, despite the apparent simplicity of the problem -- a two-dimensional two-body dynamics where both total momentum and total angular momentum are conserved -- an additional effect of cooling or heating was reported, on time scales much longer than that needed for the two atoms to oscillate. These results were obtained by numerically integrating the coupled differential equations for the internal and external degrees of freedom.

In this work, we further investigate the coupling between the dipole dynamics and the center of mass dynamics to elucidate the slow change in temperature of the system, and we study the impact of the stochastic heating due to spontaneous emission (SE). In particular, we show how friction (or anti-friction) terms appear beyond the adiabatic approximation, which explains the cooling and heating regimes. The dipole evolves on a time scale typically much shorter than the period of oscillation of the atoms center of mass in the optical potential, which allows for a multiple scale analysis. 
This purely deterministic analysis confirms that light detuned positively from the atomic transition mainly results in only metastable (heating) bound states, whereas a negative detuning rather results in stable (cooling) bound states. Yet, accounting for the stochastic heating due to spontaneous emission, one finds that the trapping potential is unable to maintain the binding forever. Just as a single particle cannot be trapped in the stationary wave created by the same beams that cools it, optical binding fails as spontaneous emission is dominated by the scattering from the light coming directly from the laser, while the optical potential results from the scattering of that laser light by one atom onto the other, so it is necessarily weaker.
As a consequence, while the presence of angular momentum in such atom pair is associated to a more efficient cooling, the lesser depth of the trapping potential makes these rotating states unstable as well.

\section{Two--atom adiabatic dynamics}

\begin{figure}
\includegraphics[width=8cm]{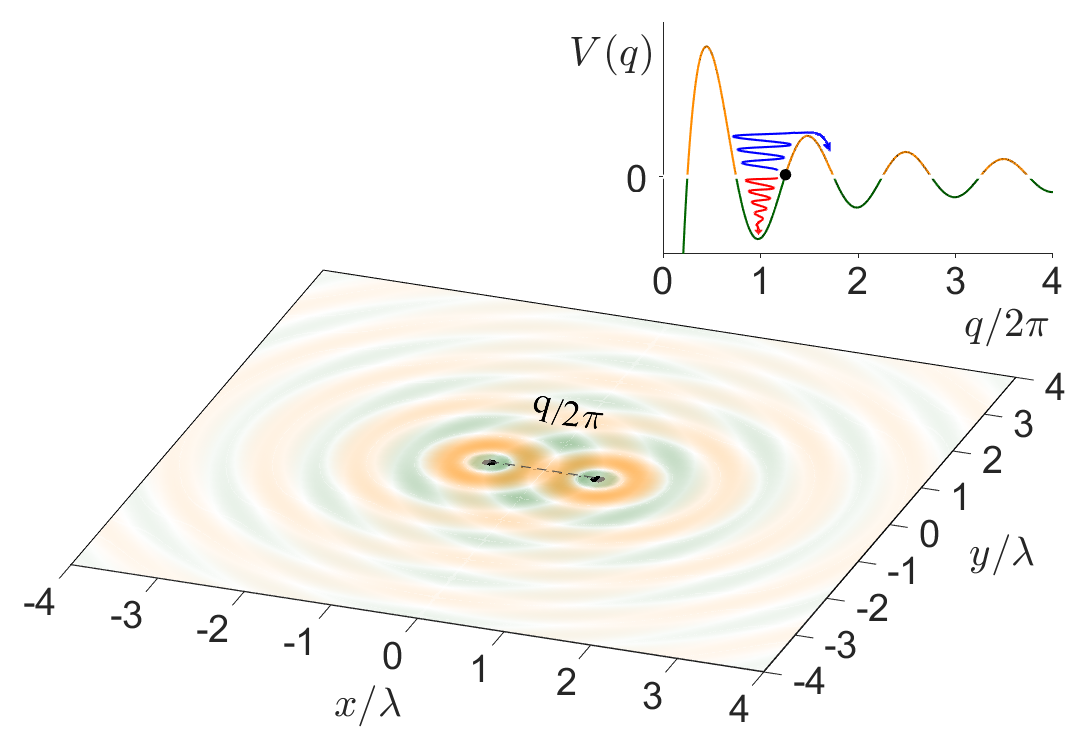}
\caption{Optical potential landscape generated by the interference between the confining laser beams (perpendicular to the plane $(x,y)$, not shown in this figure) and the radiation of the atoms. The pair of atoms is trapped in the first minimum of potential, with $|\mathbf{r}_1-\mathbf{r}_2|\approx\lambda$. The upper inset describes the profile of the self--generated potential $V(q)$ where $q=k|\mathbf{r}_1-\mathbf{r}_2|$, in absence of angular momentum.} \label{fig:scheme}
\end{figure}

Let us consider $N$ two-level atoms (polarization effects are neglected) with an atomic transition of linewidth $\Gamma$ and frequency $\omega_a$, with positions $\mathbf{r}_j$, $j=1..N$. The atoms are pumped with a monochromatic plane wave of wavevector $\mathbf{k} = k \mathbf{\hat{z}}$, detuned from the atomic transition by $\Delta = \omega - \omega_a$, and with Rabi frequency $\Omega(\mathbf{r}_j)\ll\Gamma$. Using the Markov approximation, the resonant dynamics of the atomic dipoles $\beta_j$ is given by a set of $N$ coupled equations~\citep{Lehmberg1970,Courteille2010}:
\begin{equation}
\dot\beta_j=\left(i\Delta-\frac{\Gamma}{2}\right)\beta_j-i\Omega(\mathbf{r}_j)-\frac{\Gamma}{2}\sum_{m\neq j}G_{jm}\beta_m,\label{eq:betaj0}
\end{equation}
where $G_{jm}=\exp(ik|\mathbf{r}_j-\mathbf{r}_m|)/(ik|\mathbf{r}_j-\mathbf{r}_m|)$ describes the light-mediated interaction between the dipoles. The set of equations \eqref{eq:betaj0} is linear in the dipoles $\beta_j$, so for motionless atoms most of the information on the system can be obtained from the eigenvalues and eigenvectors of the scattering matrix $G_{jm}$~\cite{Goetschy2011,Skipetrov2011,Skipetrov2014,Skipetrov2016,Guerin2017}. Neglecting the modification of the lifetime due to the atoms cooperation, the dipoles relax to equilibrium on a timescale $1/\Gamma$.
However, accounting for the optical forces resulting from the multiple light scattering leads to an intrinsically nonlinear problem, as the dynamics of the atoms center of mass couples to that of the dipoles:
\begin{equation}
m\ddot{\mathbf{r}}_j =-\hbar\Gamma\sum_{m\neq j}\mathrm{Im}\left(\nabla_{\mathbf{r}_j}G_{jm}\beta_j^*\beta_m\right).\label{eq:rj}
\end{equation}
This equation describes the average optical force between the two atoms, without accounting for the fluctuations which originate in the scattering of both laser light (spontaneous emission) and multiply scattered light (fluctuations in the dipolar force, see Sec.\ref{sec:fluct}). From now on we focus on atoms confined in a plane by counter-propagating beams, as shown in Fig.\ref{fig:scheme}. Assuming a plane wave profile for these beams, the atoms are submitted to a uniform field $\Omega$, plus the light scattered by the other atom. 
Furthermore, we restrict our analysis to pairs of atoms ($N=2$), for which the set of Eqs.(\ref{eq:betaj0}) and (\ref{eq:rj}) can be cast in the relative coordinate frame $b=(\beta_1-\beta_2)/2$, $\beta=(\beta_1+\beta_2)/2$ and $\mathbf{q}=k(\mathbf{r}_1-\mathbf{r}_2)$. In polar coordinates $\mathbf{q}=q(\cos\theta,\sin\theta)$ (where $q=kr$), one obtains~\cite{Maximo2018}
\begin{subequations}
\begin{eqnarray}
\dot b&=&-\left[1-\frac{\sin q}{q}-i\left(2\delta-\frac{\cos q}{q}\right)\right]\frac{b}{2},\label{eq:b}\\
\dot \beta&=&-\left[1+\frac{\sin q}{q}-i\left(2\delta+\frac{\cos q}{q}\right)\right]\frac{\beta}{2}-i\frac{\Omega}{\Gamma},\label{eq:beta}\\
\ddot q &=& \frac{4\omega_r}{\Gamma}\left[
\frac{4\Omega^2}{\Gamma^2}\frac{\ell^2}{q^3}-
\left(\frac{\sin q}{q}+\frac{\cos q}{q^2}\right)\left(|\beta|^2-|b|^2\right)\right],\label{eq:q}
\\ \dot{\ell}&=&0,\label{eq:l}
\end{eqnarray}
\end{subequations}
where time has been renormalized by the atomic dipole lifetime $1/\Gamma$. Here $\ell=\sqrt{\omega_r\Gamma}(L/\hbar\Omega)$, where $L=(m/2)r^2\dot\theta$ is the  total angular momentum, $\omega_r=\hbar k^2/2m$ is the recoil frequency and $\delta=\Delta/\Gamma$ the normalized detuning. Eq.\eqref{eq:l} describes the conservation of the angular momentum: Including stochastic effects such as random momentum kicks due to spontaneous emission would break this conservation law.

\begin{figure}
\includegraphics[width=8cm]{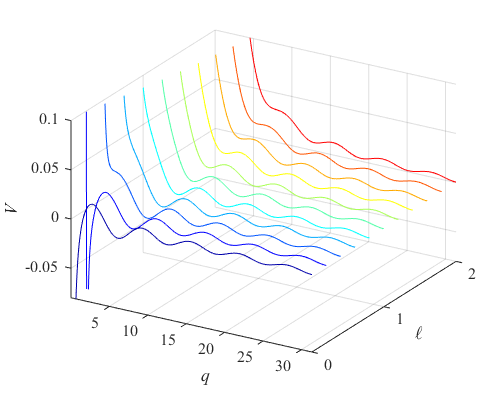}
\caption{Potential landscape $V(q)$ for different angular momenta $\ell$, for $\delta=-2$.} \label{fig:potential}
\end{figure}

Eq.\eqref{eq:b} shows that $b$ decays to zero on the dipole timescale, so the two atomic dipoles become synchronized: $\beta_1=\beta_2=\beta$. After this short transient, the equations of motion reduce to:
\begin{subequations}\label{eq:com}
\begin{eqnarray}
\dot \beta &=& -\left[1+\frac{\sin q}{q}-i\left(2\delta+\frac{\cos q}{q}\right)\right]\frac{\beta}{2}-i\frac{\Omega}{\Gamma},\label{eq:beta1}\\
\ddot q &=& \frac{4\omega_r}{\Gamma}\left[
\frac{4\Omega^2}{\Gamma^2}\frac{\ell^2}{q^3}-
\left(\frac{\sin q}{q}+\frac{\cos q}{q^2}\right)|\beta|^2\right].\label{eq:q1}
\end{eqnarray}
\end{subequations}
In order to capture the features of the short-time dynamics, we first perform the adiabatic elimination of the dipole dynamics assuming that it is synchronized with the local field. The value of $\beta$ is obtained from Eq.\eqref{eq:beta1} assuming that $\dot\beta=0$ at any time; then, inserting this value in Eq.\eqref{eq:q1} leads to:
\begin{equation}
\ddot q= \epsilon^2\left[\frac{\ell^2}{q^3}-
w(q)\right],\label{eq:q2}
\end{equation}
where we have introduced the ``small'' parameter
\begin{equation}
\epsilon=\frac{4\Omega}{\Gamma}\sqrt{\frac{\omega_r}{\Gamma}}
\end{equation}
and the function:
$$w(q)=\frac{\sin q/q+\cos q/q^2}{(1+\sin q/q)^2+(2\delta+\cos q/q)^2}.$$
Thus, in the adiabatic approximation, the dynamics of $q$ can be derived from a potential $V(q)$ given by:
\begin{equation}
V(q)=\epsilon^2 \int_q^{+\infty}\left(
 \frac{\ell^2}{q^3}-w(q)\right)dq.\label{eq:V0}
\end{equation}
The potential landscape as a function of the angular momentum is presented in Fig.~\ref{fig:potential}, where a succession of minima can be observed. For large distances $q$ between the two atoms, the potential wells become increasingly shallow as the potential decreases as $-(\cos q)/q$~\cite{Burns1989}. Furthermore, the centrifugal force opposes to the presence of low-$q$ potential minima, as can be observed for large values of the angular momentum $\ell$.
The extrema $q_n$ of this potential are given by the equation:
\begin{equation}
q_n^3 w(q_n) = \ell^2. \label{xeq}
\end{equation}
so for small angular momentum $\ell$, the stable and unstable points are found at, respectively:
\begin{subequations}\label{eq:qn}
\begin{eqnarray}
q_n^s &\approx& 2\pi n-\frac{1}{2\pi n}+\frac{\ell^2(1+4\delta^2)}{(2\pi n)^2},\label{eq:qns}
\\ q_n^u &\approx& \pi(2n+1)-\frac{1}{\pi (2n+1)}+\frac{\ell^2(1+4\delta^2)}{\pi^2 (2n+1)^2}.
\end{eqnarray}  
\end{subequations}
The potential $V$ around these points can be approximated by
\begin{equation}
V(q)\approx \epsilon^2\left[\frac{\ell^2}{2q^2}-\frac{1}{1+4\delta^2}\frac{\cos q}{q}\right].\label{eq:Veq}
\end{equation}
In particular, the potential barrier that a pair of atoms close to the point $q_n^s$ has to overcome is 
\begin{eqnarray}
U_n &=& V(q_n^u)-V(q_n^s) \nonumber
\\ &\approx& \frac{\epsilon^2}{2\pi}\frac{4n+1}{n(2n+1)}\left[\frac{1}{1+4\delta^2}-\frac{\ell^2}{4n(2n+1)}\right],\label{eq:Un}
\end{eqnarray}
which defines an admissible kinetic energy for the two particles, along the radial direction, to remain bound together.
Hence, if the pair of atoms has initially a difference of radial velocities $\delta v$, it will form a bound state provided $m(\delta v/2)^2<(\hbar\Gamma^2/4\omega_r)U_n$, or a free particles state otherwise. The system is insensitive to a velocity of the system center of mass, and difference of normal velocities correspond to the angular momentum $\ell$. Due to the integrable nature of Eq.~\eqref{eq:q2}, the bound state undergoes everlasting oscillations, with an amplitude which does not vary over time.

This long--term stability is in contrast to the results reported in Ref.~\cite{Maximo2018}, where either a slow cooling and heating of the bound system was observed by numerical integration of Eqs.(\ref{eq:com}). To explain these results, we show in the next section that the finite time needed for the dipole to equilibrate with the local field is responsible for introducing a dissipative force in Eq.\eqref{eq:q2}.

\section{Multiscale analysis\label{MSA}}
 
In general, there is a clear separation of the time scales of dipole and of the bound state vibrational mode. For example, for the Rubidium atoms probed on a MHz transition with a low pump ($\Omega\ll\Gamma$) an oscillation of the bound state spans over hundreds of dipole lifetimes \cite{Maximo2018}. More generally, one can observe from Eq.\eqref{eq:Veq} that if $\epsilon\ll1$ and $\epsilon\ell\ll1$, the vibrational mode will have a period much longer than the dipole relaxation time $1/\Gamma$.

This difference in time scales allows us to treat the finite time for the dipole equilibration as a correction to the adiabatic equation \eqref{eq:q2}. Let us introduce $g(t)=\exp[iq(t)]/[iq(t)]$ the kernel which appears in the dipole dynamics Eq.\eqref{eq:beta1}, and which varies slowly as compared to the dipole lifetime. As derived in Appendix \ref{Appendix:A}, the first correction to the adiabatic approximation reads:
\begin{equation}
\beta(t)\approx 
-\frac{2i\Omega/\Gamma}{[1-2i\delta+g(t)]}-\frac{4i\Omega}{\Gamma}\frac{\dot g(t)}{[1-2i\delta+g(t)]^3},
\label{B:beyond}
\end{equation}
where the first right-hand term corresponds to the adiabatic contribution, for which $\beta(t)$ follows instantaneously the evolution of $q(t)$. The second one describes, at first order, the delay in the dipole response to the atomic motion, and is proportional to $\dot q$. Inserting the above equation into \eqref{eq:q1} and keeping only the linear term in $\dot q$ leads to a non-conservative equation for the atoms motion:
\begin{equation}
\ddot q =-\frac{dV}{dq}-\epsilon^2 \lambda(q)\dot q,
\label{eq:nonlin}
\end{equation}
where $\lambda(q)$ is a ``friction'' coefficient which takes positive and negative value as $q$ oscillates:
\begin{widetext}
\begin{equation}
\lambda(q)=\frac{4w(q)}{\left(1+\frac{\sin q}{q}\right)^2+\left(2\delta+\frac{\cos q}{q}\right)^2}\left[\frac{\cos q}{q}-\frac{\sin q}{q^2}-2w(q)\left(1+\frac{\sin q}{q}\right)\left(2\delta+\frac{\cos q}{q}\right)\right].\label{eq:friction}
\end{equation}
\end{widetext}
From Eqs.\eqref{eq:V0} and \eqref{eq:nonlin} it becomes clear that $\dot q$ scales as $\epsilon$, so the deviation from the adiabatic dynamics of Eq.\eqref{eq:q2} occurs on a timescale $1/\epsilon$ longer than the oscillations of the bound state. The long-term consequences of the non-conservative term $\lambda(q)$ will depend on its average value over an oscillation, as we now show through a multiscale analysis.

The separation of the two timescales is realized introducing the time variables $u=\epsilon t$, associated to the oscillation of the bound state, and $v=\epsilon^2 t$, over which the dynamics drifts from its adiabatic approximation. The distance $q(u,v)$ is now considered to be a function of those two, taken to be independent variables, with the chain rule
\begin{equation}
\frac{d}{dt}=\epsilon\frac{\partial}{\partial u}+\epsilon^2\frac{\partial}{\partial v}.
\end{equation}
Applying the above rule to Eq.\eqref{eq:nonlin} leads to the multiscale equation:
\begin{eqnarray}
\frac{\partial^2 q}{\partial u^2}-\frac{\ell^2}{q^3}+w(q)&=&-2\epsilon\frac{\partial^2 q}{\partial u\partial v}-\epsilon\lambda(q)\frac{\partial q}{\partial u}\nonumber\\
&-&\epsilon^2\frac{\partial^2 q}{\partial v^2}-\epsilon^2 \lambda(q)\frac{\partial q}{\partial v}.\label{xts}
\end{eqnarray}
The separation of time scales is operated by considering the perturbation expansion $q=\sum_{n=0}^\infty \epsilon^n q_{(n)}$ which results, at the zero order in $\epsilon$, in:
\begin{eqnarray}
\frac{\partial^2 q_{(0)}}{\partial u^2}&=& \frac{\ell^2}{q_{(0)}^3} -w(q_{(0)}).\label{eq:o1}
\end{eqnarray}
It describes the adiabatic dynamics of $q_{(0)}$, i.e., it is formally equivalent to Eq.\eqref{eq:q2}. It can be associated to the potential energy $V_1=V(q_{(0)})/\epsilon^2$ from Eq.\eqref{eq:V0}, so that it admits the following energy as an integral of motion:
\begin{equation}
E(v)=\frac{1}{2}\left(\frac{\partial q_{(0)}}{\partial u}\right)^2+V_1(q_{(0)}).\label{eq:energy}
\end{equation}
This energy of the bound state varies only over the slow time scale $v$, and this drift is captured by the next order equation resulting from Eq.\eqref{xts}, which contains the non-conservative contribution:
\begin{equation}
\frac{\partial^2 q_{(1)}}{\partial u^2}+\left[\frac{3\ell^2}{q_{(0)}^4}+w'(q_{(0)})\right]q_{(1)}= -2\frac{\partial^2 q_{(0)}}{\partial u\partial v}-\lambda(q_{(0)})\frac{\partial q_{(0)}}{\partial u}.\label{eq:o2}\nonumber
\end{equation}
In order to prevent the secular growth in $q_{(1)}$, its right-hand term must vanish, a condition which reads:
\begin{equation}
\left[2\frac{\partial}{\partial v}+\lambda(q_{(0)})\right]\frac{\partial q_{(0)}}{\partial u}=0.
\end{equation}
For a bound state, the energy definition \eqref{eq:energy} provides the expression:
\begin{equation}
\frac{\partial q_{(0)}}{\partial u}=\pm \sqrt{2[E(v)-V_1(q_{(0)})]},\label{eq:dq1}
\end{equation}
which in turn leads to the equation for the evolution of the energy $E(v)$:
\begin{equation}
\frac{dE}{dv}
=-\lambda(q_{(0)})[E(v)-V_1(q_{(0)})]
+\frac{dV_1}{dq_{(0)}}\frac{\partial q_{(0)}}{\partial v}.\label{eq:dE}
\end{equation}
The slow evolution of the bound state energy is captured by integrating Eq.~\eqref{eq:dE} over a period $T$ of its oscillation:
\begin{equation}
T=2\int_{q_-}^{q_+}\frac{dq}{\sqrt{2[E(v)-V_1(q)]}},\label{period}
\end{equation}
where $q_\pm$ correspond to the extrema of the position, at which $\partial q_{(0)}/\partial u=0$. These extrema slowly change over time, so they are actually functions of $v$. 
The averaging of Eq.\eqref{eq:dE} is realized dropping its last term as it cancels over an oscillation cycle, so one obtains
\begin{equation}
\left\langle\frac{dE}{dv}\right\rangle_T
=-\frac{1}{T}\int_{q_-}^{q_+}\lambda(q)\sqrt{2[E(v)-V_1(q)]}dq.\label{eq:dE2}
\end{equation}
This equation describes the long-term evolution of the bound state energy, and predicts whether it is truly stable or only metastable.

The exact evolution of $\langle E(v)\rangle_T$ requires a numerical integration, nonetheless its behavior close to the equilibrium point $q_n^s$, given by Eq.\eqref{eq:qn}, can be captured by approximating the system as an harmonic oscillator. Introducing $\tilde{q}_n=q-q_n^s$ the relative oscillation, $\omega_n=\sqrt{V''(q_n^s)}$ its angular frequency and $\tilde{E}_n=\langle E\rangle_T-V_1(q_n^s)$ the energy relative to the equilibrium point, one can write
\begin{subequations}
\begin{eqnarray}
\langle E(v)\rangle_T &\approx& V_1(q)+\tilde{E}_n(v)-\omega_n^2\frac{\tilde{q}_n^2}{2},\label{E:exp}
\\ \lambda(q) &\approx& \lambda(q_n^s)+\lambda'(q_n)\tilde{q}_n+\lambda''(q_n^s)\frac{\tilde{q}_n^2}{2}\label{lambda:exp},
\\ q_\pm &=& q_n^s\pm \frac{\sqrt{2\tilde{E}_n(v)}}{\omega_n}\label{qpm}
\end{eqnarray}
\end{subequations}
and $T=2\pi/\omega_n$. Inserting these equations into Eq.\eqref{eq:dE2}, one finds that the linear contribution $\lambda'(q_n^s)$ of the friction term does not contribute due to the symmetry of the integral, and the remaining terms integrate as:
\begin{subequations}
\begin{eqnarray}
\frac{d\tilde{E}_n}{dv} &=& -\alpha_n\tilde{E}_n-\beta_n\tilde{E}_n^2,\label{eq:dEtilde}
\\ \alpha_n &=& \frac{\lambda(q_n^s)}{2},\label{eq:alphan}
\\ \beta_n &=& \frac{\lambda''(q_n^s)}{8\omega_n^2}.\label{eq:betan}
\end{eqnarray}
\end{subequations}
The energy $\tilde{E}_n$ is associated to the oscillations of the pair of atoms in the potential well. Due to the conservation of the angular momentum, it is naturally associated to a variation of the angular velocity as well, but it can essentially be understood as energy in the vibrational mode of the cold molecule, which can either increase (heating) or decrease (cooling) in time. 
Eq.\eqref{eq:dEtilde} describes this slow drift, over a time scale $1/\epsilon$ longer than the oscillations of the bound state, and the next section is dedicated to the different relaxation regimes.

\section{Stability of the bound states}

\subsection{Stability regions}

\begin{figure*}
\centering{\includegraphics[width=18cm]{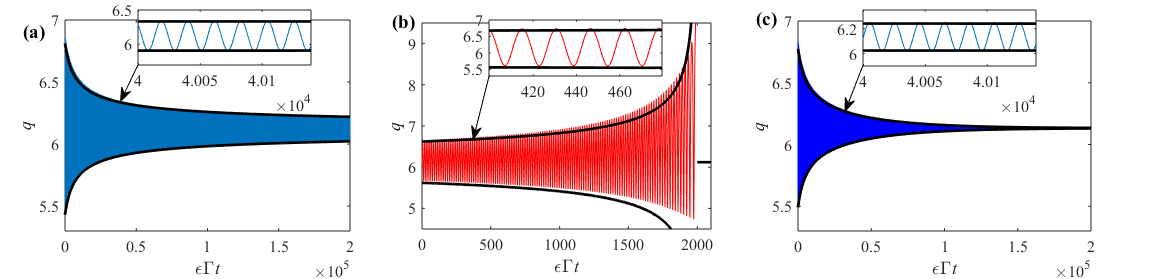}}
\caption{Dynamics of the interparticle distance $q$ for a pair of atoms (a) without angular momentum ($\ell=0$) and in the cooling regime ($\delta=-0.56$), (b) without angular momentum ($\ell=0$) and in the heating regime ($\delta=0$), and (c) with angular momentum ($\ell=0.5$) and in the cooling regime ($\delta=-0.5$). The other parameters are  $E_i=0.02$ and $\epsilon=0.1$.
The black curves correspond to the theoretical predictions of Eqs.(\ref{qpm}), (\ref{Ev:l0}) and (\ref{Ev:ln0}), where $\omega_1$, $\alpha_1$ and $\beta_1$ are given by Eqs.(\ref{omega:n})-(\ref{eq:beta}).} \label{fig:traj}
\end{figure*}

Let us first discuss the case of a bound state without angular momentum ($\ell=0$), where the two atoms oscillate along a given direction. The equilibrium condition \eqref{xeq} shows that $w(q_n^s)=0$, so the friction term \eqref{eq:friction} has no zero order contribution ($\lambda(q_n^s)=0$) and only the quadratic term in the relaxation equation \eqref{eq:dEtilde} is present. Calling $E_i=\tilde{E}_n(0)>0$ the initial energy relative to the equilibrium point $q_n^s$, and assuming that $E_i<U_n$ given by Eq.\eqref{eq:Un}, the bound state energy will drift as
\begin{equation}
\tilde{E}_n(v)=\frac{E_i}{1+\beta_n E_iv}.
\label{Ev:l0}
\end{equation}
Thus for $\beta_n>0$ the bound state will approach the equilibrium point at an algebraic speed, and the system is in a cooling regime. The time for the energy to decrease to one half of its initial value is:
\begin{equation}
\tau_n^{(1/2)}=\frac{1}{\epsilon^2\beta_n E_i}.\qquad(\ell=0)\label{t:12:lo}
\end{equation}
This behaviour is illustrated in Fig.\ref{fig:traj}(a), where the distance between a pair of atoms in the cooling regime is shown to slowly decrease over time. 

On the contrary, for $\beta_n<0$ the atomic system is heating, and the bounded pair of atoms break up as its energy reaches the potential barrier $U_n$, provided by Eq.\eqref{eq:Un}. The time for the pair of atoms to reach the escape energy is given by
\begin{equation}
\tau_n^{(esc)}=\frac{1}{\epsilon^2|\beta_n|}\left(\frac{1}{E_i}-\frac{1}{U_n}\right).\label{t:esc:l0}
\end{equation}
As depicted in Fig.\ref{fig:traj}(b), the atoms present larger and larger oscillations, until they separate and have quasi-ballistic trajectories. Finally, for $\beta_n=0$, the analysis of higher-order contributions in the friction term is necessary to determine the stability of the bound state.

In presence of angular momentum ($\ell>0$) the friction term has in general a constant contribution around the equilibrium ($\lambda(q_n^s)\neq 0$), in which case the evolution of the bound state energy reads
\begin{equation}
\tilde{E}(v)=\frac{\alpha_n E_ie^{-\alpha_n v}}{\alpha_n+\beta_nE_i\left(1-e^{-\alpha_n v}\right)}.
\label{Ev:ln0}
\end{equation}
Thus if $\alpha_n>0$ and $\alpha_n+\beta_n E_i>0$, after a transient the energy $\tilde{E}(v)$ decays exponentially fast to zero, at rate $\alpha_n$. The final bound state thus has angular momentum, but no motion in the vibrational mode, see Fig.\ref{fig:traj}(c). More generally, the half-life decay time of the energy is
\begin{equation}
\tau_n^{(1/2)}=\frac{1}{\epsilon^2\alpha_n}\ln\left[
\frac{2\alpha_n+\beta_nE_i}{\alpha_n+\beta_nE_i}\right].\label{t:12}
\end{equation}
Whereas if $\alpha_n<0$ and $\beta_n>0$, the system decreases exponentially fast, at rate $|\alpha_n|$ toward a bound state that possesses both angular momentum and energy in the vibrational mode: $\tilde{E}(\infty)=|\alpha_n|/\beta_n$. This regime sustains everlasting oscillations.

\begin{figure}
\includegraphics[width=9cm]{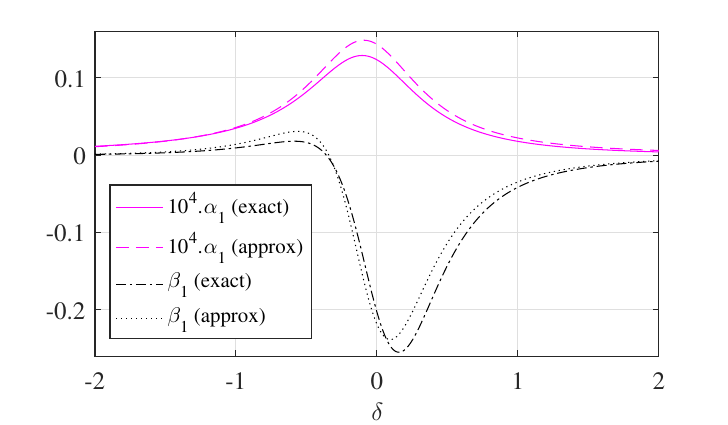}
\caption{Stability coefficients $\alpha_n$ and $\beta_n$, as calculated from Eqs.\eqref{eq:alphan}-\eqref{eq:betan} ('exact'), and from Eqs.\eqref{eq:alpha}-\eqref{eq:beta2} ('approx'). Simulations realized for $\epsilon=0.1$ and $\ell=0.1$.} \label{fig:ab}
\end{figure}

The other case, with $\alpha_n>0$ and $\beta_n<0$ such that $\alpha_n<|\beta_n|E_i$, corresponds to a bound states which is only metastable, the lifetime of which is given by
\begin{equation}
\tau_n^{(esc)}=\frac{1}{\epsilon^2\alpha_n}\ln\left[
\frac{|\beta_n|-\alpha_n/U_n}{|\beta_n|-\alpha_n/E_i}\right].\label{t:esc}
\end{equation}
Let us now provide an approximated expression of these stability parameters, by doing an expansion around the equilibrium points~\eqref{eq:qns}:
\begin{subequations}\label{eq:stabapprox}
\begin{eqnarray}
\omega_n^2 &\approx& \frac{1}{2\pi n(1+4\delta^2)},\label{omega:n}
\\ \alpha_n &\approx & \frac{\ell^2}{8(\pi n)^4(1+4\delta^2)}
\left[1-\frac{2\ell^2(\delta+1/4\pi n)}{(\pi n)^2}\right],\label{eq:alpha}
\\ \beta_n  &\approx & -\frac{2}{\pi n(1+4\delta^2)^2}\left[
\delta+\frac{1+\delta^2}{\pi n}\right].\label{eq:beta2}
\end{eqnarray}
\end{subequations}
Let us first discuss the case without angular momentum, where only the $\beta_n$ coefficient is relevant (see Eq.\eqref{Ev:l0}). In this case, under the condition
\begin{equation}
-\sqrt{\left(\frac{n\pi}{2}\right)^2-1}-\frac{n\pi}{2}\leq \delta \leq \sqrt{\left(\frac{n\pi}{2}\right)^2-1}-\frac{n\pi}{2},
\end{equation}
the $\beta_n$ coefficient is positive and the bound states are truly stable. Otherwise, $\beta_n$ is negative and the bound states are only metastable. The behavior of $\beta_n$ as a function of the detuning is illustrated in Fig.\ref{fig:ab}, where a range of negative detuning allows for stable bound states. 

In presence of a small angular momentum (that is, such that $\alpha_n$ is positive), the system is stable over a larger range of detuning, since $\beta_n>-\alpha_n/E_i$ is now a sufficient condition to reach a cooling regime.

Note that while Eq.\eqref{eq:alpha} suggests that $\alpha_n$ becomes negative for large values of angular momentum, the approximated expressions \eqref{eq:stabapprox} lose their validity, and the increase of $\ell$ actually suppresses successively the potential minima that are responsible for the bound states (see Fig.\ref{fig:potential}). A more detailed study of the high-$\ell$ regime will require different approximations than the ones performed here.

\subsection{Cooling and heating time}

\begin{figure}
\centering{\includegraphics[width=9cm]{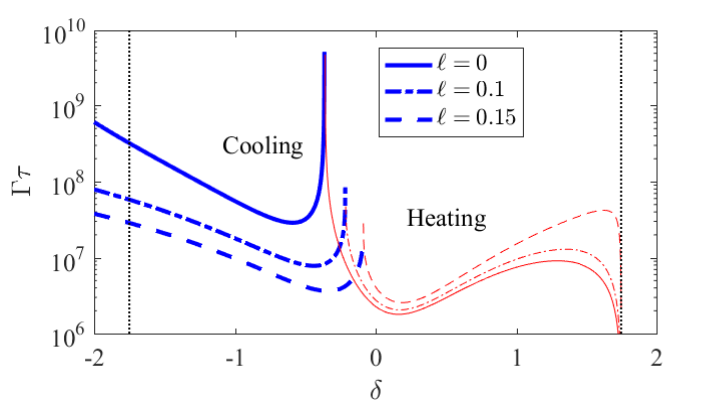}}
\caption{Cooling (thick blue lines) and heating (thin red lines) time as a function of the detuning $\delta$, for different values of the angular momentum $\ell$, for $\epsilon=0.1$ and $E_i=2.10^{-4}$. Both times present a divergence at the critical detuning where the long-term stability of the bound state changes. The vertical black dotted lines correspond to the stability threshold defined by $E_i=U_n$.}. \label{fig:tau_12}
\end{figure}

Let us first comment that the energy in the vibrational mode $\tilde{E}$ is a function of $v$, i.e., it scales with $1/\epsilon^2\sim \Gamma^3/(\Omega^2\omega_r)$. So $\epsilon$ is the fundamental parameter to control the time scales over which cooling and heating act.
Then, a numerical study of the heating and cooling times reveal that it strongly depends on the detuning, see Fig.\ref{fig:tau_12} for examples of this dependence for different values of the angular momentum. First, the heating time presents a minimum (which means the heating rate is maximum) very close to resonance ($\delta\approx 0.15\Gamma$); this is somehow expected from scattering of light very close to the atomic resonance, where the radiation pressure force dominates over the dipolar force. Instead, the cooling is most efficient for light slightly detuned to the red, with a maximum that depends significantly on the angular momentum. In both heating and cooling regimes, the rates decrease going farther away from resonance, where light-atom coupling is less efficient. For a given $\epsilon$ and initial energy $E_i$, the barrier potential $U_n$ of the bound state decreases with the detuning (see Eq.\eqref{eq:Un}), so there is no more bound state at large detuning (see vertical dotted lines in Fig.\ref{fig:tau_12}).

Interestingly, the heating rate is not very sensitive to the angular momentum, but the cooling rate is. From $\ell=0$ to $\ell=0.15$, a factor $\sim10$ is gained on the cooling rate of the bound state. This highlights that the angular momentum of the system increases the stability of the system, possibly countering other heating effects.

\begin{figure}
\centering{\includegraphics[width=8.5cm]{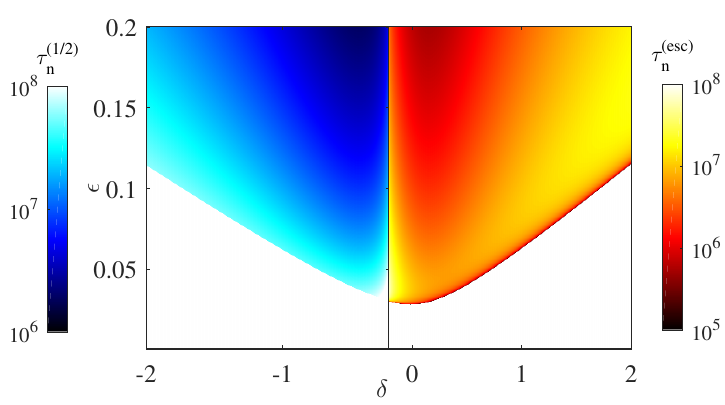}}
\caption{Heating and cooling times as a function of the detuning $\delta$ and the parameter $\epsilon$, for $\ell=0.1$. The negative detuning part ($\delta\lesssim-0.21$) corresponds to the cooling regime (blue colormap), whereas the positive detuning part ($\delta\gtrsim -0.21$) stands for the heating regime. The black vertical line marks the separation between the two regimes, and the white area corresponds to unbound states ($E_i>U_n$). Simulations realized for $E_i=2.10^{-4}$ and $n=1$, using Eqs.(\ref{t:12}--\ref{eq:stabapprox}).} \label{tau_map}
\end{figure}

A stability diagram is presented in Fig.\ref{tau_map} for $\ell=0.1$, showing the heating and cooling times as a function of the detuning and of the parameter $\epsilon$. A larger pump strength enhances in atom-light coupling, and thus results in a higher rate of change in the energy of the bound state, just like working close to resonance.

\section{Impact of the fluctuations\label{sec:fluct}}

\begin{figure}
\centering{\includegraphics[width=9.5cm]{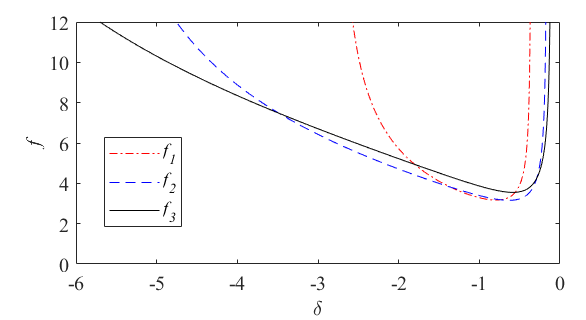}}
\caption{Amplitude of the equilibrum energy $f(\delta)$, in units of $\hbar\Omega$, for different detuning $\delta$, as predicted by the stochastic contribution and in the range where $\beta_n$ is positive (cooling regime).} \label{fd}
\end{figure}

The analysis up to now was purely deterministic, neglecting the effect of the fluctuations due to spontaneous emission as the atoms interact with the incident lasers, and with their mutual radiation. The atoms receive a random momentum kick $\delta \mathbf{p}=\hbar\mathbf{k}$ which introduces a stochastic contribution both in the radial and in the angular directions. Each scattering event results in an average increase of the associated energy of $\delta E_{\mathrm{recoil}}=\hbar\omega_r/2$. Focusing at first on spontaneous emission from the driving of the confining lasers, the heating energy rate is proportional to the scattering rate:
\begin{equation}
\left(\frac{\delta E}{\delta t}\right)_{\mathrm{SE}}=\frac{2\hbar\omega_r}{\Gamma}\frac{\Omega^2}{1+4\delta^2}.\label{spont}
\end{equation}
Adding this term to the equation for the scaled average energy $\tilde E_n$ for the radial energy in Eq.(\ref{eq:dEtilde}), using the relations $\tilde E=(4\omega_r/\Gamma\epsilon^2)(E/\hbar\Gamma)$ and $v=\epsilon^2\Gamma t$, one obtains
\begin{equation}
\frac{d\tilde{E}_n}{dt} = -\epsilon^2\Gamma(\alpha_n\tilde{E}_n+\beta_n\tilde{E}_n^2)+\frac{\omega_r^2}{2(1+4\delta^2)}.
\end{equation}
The steady-state solution is thus given by
\begin{equation}
\tilde E_n^{\infty}=\frac{1}{2\beta_n}\left(\sqrt{\alpha_n^2+4\beta_n C}-\alpha_n\right)
\end{equation}
where $C=(\omega_r/2\Gamma\epsilon^2)/(1+4\delta^2)$. Since $4\beta_nC\gg\alpha_n^2$, $\tilde E_n^{\infty}\approx\sqrt{C/\beta_n},$
which, in physical units, reads
\begin{equation}
E^{\infty}_n\approx f_n(\delta)\hbar\Omega
\end{equation}
with
\begin{equation}
f_n(\delta)=\frac{1}{\sqrt{2\beta_n(1+4\delta^2)}}.
\end{equation}
The function $f_n(\delta)$ is plotted in Fig.\ref{fd} for the values where $\beta_n$ is positive (cooling regime for the deterministic dynamics), as a function of the detuning $\delta$ and for $n=1,\ 2,\ 3$. It reaches a minimum around $\delta=-3/4$, close to the value at which the cooling term $\beta_n$ is maximum, and the achieved steady-state energy is $E^{\infty}_n\approx 2\hbar\Omega$.

The fact that the limit temperature is proportional to $\hbar\Omega$ is rather surprising, as compared to the ``standard'' limit of laser cooling of $\sim \hbar\Gamma$~\cite{Letokhov1976}. However, a similar temperature  can be identified for a single two-level atom confined in a standing wave. Let us shortly review this situation: A standing wave along $z$ with $\Omega(z)=\Omega_0\cos(kz)$ produces  a force along the $z$-axis
\begin{equation}
F_z=\hbar k\frac{\Omega_0^2}{\Gamma}\frac{2\delta}{1+4\delta^2}
\left[\sin(kz)\cos(kz)+\frac{\sin^2(kz)}{1+4\delta^2}\frac{kv_z}{\Gamma}\right].
\end{equation}
When averaged over a spatial period $\lambda/2$ this force reduces to the usual viscous force $F_z=-\alpha v_z$, with $\alpha=4\hbar k^2(\Omega_0/\Gamma)^2[-2\delta/(1+4\delta^2)^2]$. If instead the atom is near the potential minimum at $z=0$, with a kinetic energy smaller than the trapping energy $\hbar\Gamma(\Omega_0/\Gamma)^2[-\delta/(1+4\delta^2)]$, the force can be locally expanded as
\begin{equation}
F_z\approx\hbar k^2\frac{\Omega_0^2}{\Gamma}\frac{2\delta}{1+4\delta^2}
\left(z+\frac{(kz)^2}{1+4\delta^2}\frac{v_z}{\Gamma}\right).
\end{equation}
For $\delta<0$ the atom is trapped by the dipole force and cooled by a force which is linear in the velocity and quadratic in the position. When averaged over the oscillating motion, a multiscale analysis similar to that performed in Sec.III leads to the following equation for the energy:
\begin{equation}
\frac{dE_z}{dt}=-\frac{2\omega_r}{1+4\delta^2}\frac{E_z^2}{\hbar\Gamma}+\frac{4}{3}\frac{\hbar\omega_r}{\Gamma}\frac{\Omega_0^2}{1+4\delta^2},
\end{equation}
from which an equilibrium energy $E_z^{\infty}=\sqrt{3/2}\hbar\Omega_0$ can be deduced. Hence, single atom  cooling {\it in a standing wave} also presents a limit temperature $\propto\hbar\Omega_0$ for low-energy initial states, in addition to the usual Doppler limit $\hbar\Gamma$.

The trick is that the linear regime assumption ($s=2\Omega_0^2/(\Gamma^2+4\Delta^2)\ll 1$) underlying the classical treatment of the atom dynamics is incompatible with the requirement of a trapping potential deeper that the equilibrium energy. Indeed the ratio between the trapping potential depth and the equilibrium energy is $\sqrt{s}(-\delta/\sqrt{1+4\delta^2})$, where the latter function of $\delta$ tops at $1/2$, so the confinement cannot be achieved at equilibrium.

In the case of optically bound pair of atoms, the ratio is even worse as SE relies on the incident laser, while the trapping potential requires an additional scattering event from the atoms. More specifically, the ratio between the trapping potential depth and the equilibrium energy is $\sim\sqrt{s/(1+4\delta^2)}$. Thus, radial confinement of the pair cannot be achieved without any additional cooling mechanism.

As for the rotational degree of freedom, the stochastic contribution leads to a pure diffusive behaviour of the angular momentum $L$, as the deterministic dynamics preserves it. The diffusion makes the transverse energy grow as $\langle L^2/mr^2\rangle\sim \hbar\omega_r s\Gamma t$.
Furthermore, the rotational motion of the molecule decreases its radial potential barrier (see Eq.\ref{eq:Un}), i.e., it makes the system even less stable. Hence, a cooling mechanism active on the angular motion of the molecules will be necessary to achieve optical binding with cold atoms.

Let us comment that another heating mechanism has been identified in Ref.\cite{Zhu2016}, which corresponds to momentum diffusion from radiative interaction, i.e., fluctuations in the dipolar force (which is here responsible for the OB). In the case of the pair of atoms, the heating rate reads
\begin{equation}
\left(\frac{\delta E}{\delta t}\right)_{\mathrm{rad}}\sim\frac{\hbar\omega_r}{\Gamma}\frac{\Omega^2}{1+4\delta^2}\overline{\nabla_q^2\left(\frac{\sin q}{q}\right)},
\end{equation}
where the bar refers to an average over the oscillation period. Nevertheless, close to the equilibrium position $q^s_n$, $\overline{\nabla_q^2(\sin q/q)}\approx 0.03$, so it only represents a correction of a few percent to the contribution of the SE from the driving laser \eqref{spont}.

\section{Discussion and conclusions}

To summarize, we have first shown that the optical binding of two atoms in the vacuum and confined in a plane, is affected by a {\it deterministic} non--conservative force able to cool or heat the system. This force arises from the non--adiabatic reaction of the atomic dipole to the change of field as the distance between the atoms change. This force is strongly position--dependent but, when averaged over an oscillation of the pair of atoms, it effectively results in a slow heating or cooling of the system. It may thus either lead the atom to escape the influence of each other, typically for positive detuning, or rather drive them toward the local potential minimum, in general for negative detuning.

In particular, the specificity of the cooling associated to the angular momentum can be better understood by analyzing further Eq.\eqref{eq:dEtilde}: The $\beta_n$ coefficient, which does not involve angular momentum at first order, is associated to a quadratic dependence in $\tilde{E}$, so it is efficient only when the system is significantly afar from the stable point. On the contrary, the $\alpha_n$ term, which scales directly with $\ell^2$, appears in a term linear with $\tilde{E}$. Hence, it acts as a ``friction'' term, and is most efficient at keeping the system very close to the equilibrium point.

Nevertheless, the effect of the stochastic heating due to spontaneous emission appears to be stronger than the confining potential that gives rises to the optical binding. Both the stretching vibrational mode and the rotational degree of freedom turn out to be ultimately dominated by diffusion effects, so the bound states are not truly bound.


The lack of stability of the OB configurations for pairs of atoms calls for alternative ways to achieve the binding. In this respect, collective effects in larger atomic systems may be a promising candidate, as the cooperative emission (such as superradiance) is an efficient mechanism for self-organization in one dimension~ \cite{Bonifacio1994,Slama2007}. As for two-dimensional systems, 
crystallization is expected to occur, thanks to the optical potential generated on each atom by its neighbours~\cite{Ng2005}. In this case many--atom effects may significantly alter the cooling properties of the system, as collective oscillation modes arise. In this context, the angular momentum may provide an extra degree of freedom to tune the stability properties of the system, but also to modify the spatial period of the crystal, and possibly its lattice structure.

\section{Acknowledgements}

This work was performed in the framework of the European Training Network ColOpt, which is funded by the European Union (EU) Horizon 2020 programme under the Marie  Sklodowska-Curie action, grant agreement 721465. RB
hold grants from S\~ao Paulo Research Foundation (FAPESP), grant 2014/01491-0. We acknowledge fruitful discussions with R. Kaiser and C.E. Maximo.

\begin{widetext}
\appendix

\section{Analysis of the adiabatic approximation\label{Appendix:A}}
In order to discuss the adiabatic approximation, let's integrate Eq.(\ref{eq:beta1}) from $0$ to $t$ with $\beta(0)=0$:
\begin{equation}
\beta(t)=-i\frac{\Omega}{\Gamma}\int_0^t dt'\exp
\left\{-\frac{1}{2}(1-2i\delta)t'-\frac{1}{2}\int_{0}^{t'}g(t-t'+t'')dt''\right\}\label{B:Int}
\end{equation}
where
\begin{equation}
g(t)=\frac{\exp[iq(t)]}{iq(t)}
\end{equation}
Let assume that $g(t)$ varies slowly with respect to the term $(1-2i\delta)t'$. However, we consider the first order deviation of $g(t)$, in order to go beyond the usual adiabatic approximation, expanding $g(t-t'+t'')$ in the integral of Eq.\ref{B:Int}) up to the first order in its Taylor series:
\begin{equation}
\int_{0}^{t'}g(t-t'+t'')dt''\approx g(t)t'-\dot g(t)\int_{0}^{t'}(t'-t'')dt''=g(t)t'-\frac{1}{2}\dot g(t)\, t'^2.
\label{int:g}
\end{equation}
The first term of Eq.(\ref{int:g}) corresponds to the usual adiabatic approximation, whereas the second term takes into account of the slow variation of $g$ due to the atomic motion in the confining potential. Since $g$ depends on the relative atomic position $q(t)$, then $\dot g$ is proportional to the relative atomic velocity.

Once inserted Eq.(\ref{int:g}) in Eq.(\ref{B:Int}), it gives:
\begin{equation}
\beta(t)\approx-i\frac{\Omega}{\Gamma}\int_0^\infty dt'\exp
\left\{-\frac{1}{2}[1-2i\delta+g(t)]t'+\frac{1}{4}\dot g(t)\, t'^2\right\},\label{B:diff}
\end{equation}
where we have extended the integration upper limit to infinity, neglecting in this way the short initial transient.
By expanding the small term proportional to $\dot g(t)$ at the first order:
\begin{eqnarray}
\beta(t)&= &-i\frac{\Omega}{\Gamma}\int_0^\infty dt'\exp
\left\{-\frac{1}{2}[1-2i\delta+g(t)]t'\right\}\left[1+\frac{1}{4}\dot g(t)\, t'^2+\dots\right]\nonumber\\
&\approx &
-i\frac{2\Omega}{\Gamma}\frac{1}{1-2i\delta+g(t)}
\left\{1+\frac{2\dot g(t)}{[1-2i\delta+g(t)]^2}\right\}
\label{B:diff:2}
\end{eqnarray}
The first term of Eq.(\ref{B:diff:2}) is the usual adiabatic approximation, whereas the second term correspond to the correction due to the atomic displacement. It is similar to the Doppler effect in the optical molasses, with the difference that here the atomic displacement is not due to the thermal motion, but to the oscillation in the optical binding potential. Also we can say that in general this velocity-dependent force is due to the cooperative decay and light shift, depending on the distance between the atoms and induced by the laser. From Eq.(\ref{B:diff:2}), we obtain
\begin{eqnarray}
|\beta(t)|^2 &=&
\frac{4\Omega^2}{\Gamma^2}
\frac{1}{D(q)}+
\frac{16\Omega^2}{\Gamma^2}\frac{1}{D^3(q)}\left[
\mathrm{Re}\dot g(t)D(q)-2\mathrm{Im}\dot g(t)(1+\sin q/q)(2\delta+\cos q/q)
\right].\label{BB}
\end{eqnarray}
where
\begin{eqnarray}
\mathrm{Re}\dot g(t)&=&\frac{d}{dq}\left(\frac{\sin q}{q}\right)\dot q=\left(\frac{\cos q}{q}-\frac{\sin q}{q^2}\right)\dot q\label{regdot}\\
\mathrm{Im}\dot g(t)&=&-\frac{d}{dq}\left(\frac{\cos q}{q}\right)\dot q
=
\left(\frac{\sin q}{q}+\frac{\cos q}{q^2}\right)\dot q.\label{imgdot}
\end{eqnarray}
and $D(q)=(1+\sin q/q)^2+(2\delta+\cos q/q)^2$. Inserting Eqs.(\ref{BB})-(\ref{imgdot}) in the force equation (\ref{eq:q1}), we obtain
\begin{eqnarray}
\ddot q &=& \frac{16\omega_r\Omega^2}{\Gamma^3}\left[
\frac{\ell^2}{q^3}-w(q)-\lambda(q)\dot q\right],\label{eq:qfric}
\end{eqnarray}
where
\begin{eqnarray}
w(q)&=& \frac{1}{D(q)}
\left(\frac{\sin q}{q}+\frac{\cos q}{q^2}\right)\\
\lambda(q)&=&-\frac{4w(q)}{D(q)}\left[\frac{\cos q}{q}-\frac{\sin q}{q^2}-2w(q)\left(1+\frac{\sin q}{q}\right)\left(2\delta+\frac{\cos q}{q}\right)\right].
\end{eqnarray}

\end{widetext}

\bibliography{./../../../Biblio/BiblioCollectiveScattering}

\begin{thebibliography}{27}%
\makeatletter
\providecommand \@ifxundefined [1]{%
 \@ifx{#1\undefined}
}%
\providecommand \@ifnum [1]{%
 \ifnum #1\expandafter \@firstoftwo
 \else \expandafter \@secondoftwo
 \fi
}%
\providecommand \@ifx [1]{%
 \ifx #1\expandafter \@firstoftwo
 \else \expandafter \@secondoftwo
 \fi
}%
\providecommand \natexlab [1]{#1}%
\providecommand \enquote  [1]{``#1''}%
\providecommand \bibnamefont  [1]{#1}%
\providecommand \bibfnamefont [1]{#1}%
\providecommand \citenamefont [1]{#1}%
\providecommand \href@noop [0]{\@secondoftwo}%
\providecommand \href [0]{\begingroup \@sanitize@url \@href}%
\providecommand \@href[1]{\@@startlink{#1}\@@href}%
\providecommand \@@href[1]{\endgroup#1\@@endlink}%
\providecommand \@sanitize@url [0]{\catcode `\\12\catcode `\$12\catcode
  `\&12\catcode `\#12\catcode `\^12\catcode `\_12\catcode `\%12\relax}%
\providecommand \@@startlink[1]{}%
\providecommand \@@endlink[0]{}%
\providecommand \url  [0]{\begingroup\@sanitize@url \@url }%
\providecommand \@url [1]{\endgroup\@href {#1}{\urlprefix }}%
\providecommand \urlprefix  [0]{URL }%
\providecommand \Eprint [0]{\href }%
\providecommand \doibase [0]{http://dx.doi.org/}%
\providecommand \selectlanguage [0]{\@gobble}%
\providecommand \bibinfo  [0]{\@secondoftwo}%
\providecommand \bibfield  [0]{\@secondoftwo}%
\providecommand \translation [1]{[#1]}%
\providecommand \BibitemOpen [0]{}%
\providecommand \bibitemStop [0]{}%
\providecommand \bibitemNoStop [0]{.\EOS\space}%
\providecommand \EOS [0]{\spacefactor3000\relax}%
\providecommand \BibitemShut  [1]{\csname bibitem#1\endcsname}%
\let\auto@bib@innerbib\@empty
\bibitem [{\citenamefont {Phillips}(1998)}]{Phillips1998}%
  \BibitemOpen
  \bibfield  {author} {\bibinfo {author} {\bibfnamefont {William~D.}\
  \bibnamefont {Phillips}},\ }\bibfield  {title} {\enquote {\bibinfo {title}
  {Nobel lecture: Laser cooling and trapping of neutral atoms},}\ }\href
  {\doibase 10.1103/revmodphys.70.721} {\bibfield  {journal} {\bibinfo
  {journal} {Reviews of Modern Physics}\ }\textbf {\bibinfo {volume} {70}},\
  \bibinfo {pages} {721--741} (\bibinfo {year} {1998})}\BibitemShut {NoStop}%
\bibitem [{\citenamefont {Anderson}\ \emph {et~al.}(1995)\citenamefont
  {Anderson}, \citenamefont {Ensher}, \citenamefont {Matthews}, \citenamefont
  {Wieman},\ and\ \citenamefont {Cornell}}]{Anderson1995}%
  \BibitemOpen
  \bibfield  {author} {\bibinfo {author} {\bibfnamefont {M.~H.}\ \bibnamefont
  {Anderson}}, \bibinfo {author} {\bibfnamefont {J.~R.}\ \bibnamefont
  {Ensher}}, \bibinfo {author} {\bibfnamefont {M.~R.}\ \bibnamefont
  {Matthews}}, \bibinfo {author} {\bibfnamefont {C.~E.}\ \bibnamefont
  {Wieman}}, \ and\ \bibinfo {author} {\bibfnamefont {E.~A.}\ \bibnamefont
  {Cornell}},\ }\bibfield  {title} {\enquote {\bibinfo {title} {Observation of
  bose-einstein condensation in a dilute atomic vapor},}\ }\href {\doibase
  10.1126/science.269.5221.198} {\bibfield  {journal} {\bibinfo  {journal}
  {Science}\ }\textbf {\bibinfo {volume} {269}},\ \bibinfo {pages} {198--201}
  (\bibinfo {year} {1995})}\BibitemShut {NoStop}%
\bibitem [{\citenamefont {Greiner}\ \emph {et~al.}(2002)\citenamefont
  {Greiner}, \citenamefont {Mandel}, \citenamefont {Esslinger}, \citenamefont
  {H\"ansch},\ and\ \citenamefont {Bloch}}]{Greiner2002}%
  \BibitemOpen
  \bibfield  {author} {\bibinfo {author} {\bibfnamefont {M.}~\bibnamefont
  {Greiner}}, \bibinfo {author} {\bibfnamefont {O.}~\bibnamefont {Mandel}},
  \bibinfo {author} {\bibfnamefont {T.}~\bibnamefont {Esslinger}}, \bibinfo
  {author} {\bibfnamefont {T.W.}\ \bibnamefont {H\"ansch}}, \ and\ \bibinfo
  {author} {\bibfnamefont {I.}~\bibnamefont {Bloch}},\ }\bibfield  {title}
  {\enquote {\bibinfo {title} {Quantum phase transition from a superfluid to a
  mott insulator in a gas of ultracold atoms},}\ }\href@noop {} {\bibfield
  {journal} {\bibinfo  {journal} {Nature}\ }\textbf {\bibinfo {volume} {415}},\
  \bibinfo {pages} {39--44} (\bibinfo {year} {2002})}\BibitemShut {NoStop}%
\bibitem [{\citenamefont {Myatt}\ \emph {et~al.}(1997)\citenamefont {Myatt},
  \citenamefont {Burt}, \citenamefont {Ghrist}, \citenamefont {Cornell},\ and\
  \citenamefont {Wieman}}]{Myatt1997}%
  \BibitemOpen
  \bibfield  {author} {\bibinfo {author} {\bibfnamefont {C.~J.}\ \bibnamefont
  {Myatt}}, \bibinfo {author} {\bibfnamefont {E.~A.}\ \bibnamefont {Burt}},
  \bibinfo {author} {\bibfnamefont {R.~W.}\ \bibnamefont {Ghrist}}, \bibinfo
  {author} {\bibfnamefont {E.~A.}\ \bibnamefont {Cornell}}, \ and\ \bibinfo
  {author} {\bibfnamefont {C.~E.}\ \bibnamefont {Wieman}},\ }\bibfield  {title}
  {\enquote {\bibinfo {title} {Production of two overlapping bose-einstein
  condensates by sympathetic cooling},}\ }\href {\doibase
  10.1103/physrevlett.78.586} {\bibfield  {journal} {\bibinfo  {journal}
  {Physical Review Letters}\ }\textbf {\bibinfo {volume} {78}},\ \bibinfo
  {pages} {586--589} (\bibinfo {year} {1997})}\BibitemShut {NoStop}%
\bibitem [{\citenamefont {Labeyrie}\ \emph {et~al.}(2014)\citenamefont
  {Labeyrie}, \citenamefont {Tesio}, \citenamefont {Gomes}, \citenamefont
  {Oppo}, \citenamefont {Firth}, \citenamefont {Robb}, \citenamefont {Arnold},
  \citenamefont {Kaiser},\ and\ \citenamefont {Ackemann}}]{Labeyrie2014}%
  \BibitemOpen
  \bibfield  {author} {\bibinfo {author} {\bibfnamefont {G.}~\bibnamefont
  {Labeyrie}}, \bibinfo {author} {\bibfnamefont {E.}~\bibnamefont {Tesio}},
  \bibinfo {author} {\bibfnamefont {P.~M.}\ \bibnamefont {Gomes}}, \bibinfo
  {author} {\bibfnamefont {G.-L.}\ \bibnamefont {Oppo}}, \bibinfo {author}
  {\bibfnamefont {W.~J.}\ \bibnamefont {Firth}}, \bibinfo {author}
  {\bibfnamefont {G.~R.~M.}\ \bibnamefont {Robb}}, \bibinfo {author}
  {\bibfnamefont {A.~S.}\ \bibnamefont {Arnold}}, \bibinfo {author}
  {\bibfnamefont {R.}~\bibnamefont {Kaiser}}, \ and\ \bibinfo {author}
  {\bibfnamefont {T.}~\bibnamefont {Ackemann}},\ }\bibfield  {title} {\enquote
  {\bibinfo {title} {Optomechanical self-structuring in a cold atomic gas},}\
  }\href {\doibase 10.1038/nphoton.2014.52} {\bibfield  {journal} {\bibinfo
  {journal} {Nature Photonics}\ }\textbf {\bibinfo {volume} {8}},\ \bibinfo
  {pages} {321--325} (\bibinfo {year} {2014})}\BibitemShut {NoStop}%
\bibitem [{\citenamefont {Bonifacio}\ and\ \citenamefont
  {Salvo}(1994)}]{Bonifacio1994}%
  \BibitemOpen
  \bibfield  {author} {\bibinfo {author} {\bibfnamefont {R.}~\bibnamefont
  {Bonifacio}}\ and\ \bibinfo {author} {\bibfnamefont {L.~De}\ \bibnamefont
  {Salvo}},\ }\bibfield  {title} {\enquote {\bibinfo {title} {Collective atomic
  recoil laser ({CARL}) optical gain without inversion by collective atomic
  recoil and self-bunching of two-level atoms},}\ }\href {\doibase
  10.1016/0168-9002(94)90382-4} {\bibfield  {journal} {\bibinfo  {journal}
  {Nuclear Instruments and Methods in Physics Research Section A: Accelerators,
  Spectrometers, Detectors and Associated Equipment}\ }\textbf {\bibinfo
  {volume} {341}},\ \bibinfo {pages} {360--362} (\bibinfo {year}
  {1994})}\BibitemShut {NoStop}%
\bibitem [{\citenamefont {Slama}\ \emph {et~al.}(2007)\citenamefont {Slama},
  \citenamefont {Bux}, \citenamefont {Krenz}, \citenamefont {Zimmermann},\ and\
  \citenamefont {Courteille}}]{Slama2007}%
  \BibitemOpen
  \bibfield  {author} {\bibinfo {author} {\bibfnamefont {S.}~\bibnamefont
  {Slama}}, \bibinfo {author} {\bibfnamefont {S.}~\bibnamefont {Bux}}, \bibinfo
  {author} {\bibfnamefont {G.}~\bibnamefont {Krenz}}, \bibinfo {author}
  {\bibfnamefont {C.}~\bibnamefont {Zimmermann}}, \ and\ \bibinfo {author}
  {\bibfnamefont {Ph.~W.}\ \bibnamefont {Courteille}},\ }\bibfield  {title}
  {\enquote {\bibinfo {title} {Superradiant rayleigh scattering and collective
  atomic recoil lasing in a ring cavity},}\ }\href {\doibase
  10.1103/physrevlett.98.053603} {\bibfield  {journal} {\bibinfo  {journal}
  {Physical Review Letters}\ }\textbf {\bibinfo {volume} {98}} (\bibinfo {year}
  {2007}),\ 10.1103/physrevlett.98.053603}\BibitemShut {NoStop}%
\bibitem [{\citenamefont {Ritsch}\ \emph {et~al.}(2013)\citenamefont {Ritsch},
  \citenamefont {Domokos}, \citenamefont {Brennecke},\ and\ \citenamefont
  {Esslinger}}]{Ritsch2013}%
  \BibitemOpen
  \bibfield  {author} {\bibinfo {author} {\bibfnamefont {Helmut}\ \bibnamefont
  {Ritsch}}, \bibinfo {author} {\bibfnamefont {Peter}\ \bibnamefont {Domokos}},
  \bibinfo {author} {\bibfnamefont {Ferdinand}\ \bibnamefont {Brennecke}}, \
  and\ \bibinfo {author} {\bibfnamefont {Tilman}\ \bibnamefont {Esslinger}},\
  }\bibfield  {title} {\enquote {\bibinfo {title} {Cold atoms in
  cavity-generated dynamical optical potentials},}\ }\href {\doibase
  10.1103/revmodphys.85.553} {\bibfield  {journal} {\bibinfo  {journal}
  {Reviews of Modern Physics}\ }\textbf {\bibinfo {volume} {85}},\ \bibinfo
  {pages} {553--601} (\bibinfo {year} {2013})}\BibitemShut {NoStop}%
\bibitem [{\citenamefont {M{\'{a}}ximo}\ \emph {et~al.}(2018)\citenamefont
  {M{\'{a}}ximo}, \citenamefont {Bachelard},\ and\ \citenamefont
  {Kaiser}}]{Maximo2018}%
  \BibitemOpen
  \bibfield  {author} {\bibinfo {author} {\bibfnamefont {C.~E.}\ \bibnamefont
  {M{\'{a}}ximo}}, \bibinfo {author} {\bibfnamefont {R.}~\bibnamefont
  {Bachelard}}, \ and\ \bibinfo {author} {\bibfnamefont {R.}~\bibnamefont
  {Kaiser}},\ }\bibfield  {title} {\enquote {\bibinfo {title} {Optical binding
  with cold atoms},}\ }\href {\doibase 10.1103/physreva.97.043845} {\bibfield
  {journal} {\bibinfo  {journal} {Physical Review A}\ }\textbf {\bibinfo
  {volume} {97}} (\bibinfo {year} {2018}),\
  10.1103/physreva.97.043845}\BibitemShut {NoStop}%
\bibitem [{\citenamefont {Burns}\ \emph {et~al.}(1989)\citenamefont {Burns},
  \citenamefont {Fournier},\ and\ \citenamefont {Golovchenko}}]{Burns1989}%
  \BibitemOpen
  \bibfield  {author} {\bibinfo {author} {\bibfnamefont {Michael~M.}\
  \bibnamefont {Burns}}, \bibinfo {author} {\bibfnamefont {Jean-Marc}\
  \bibnamefont {Fournier}}, \ and\ \bibinfo {author} {\bibfnamefont {Jene~A.}\
  \bibnamefont {Golovchenko}},\ }\bibfield  {title} {\enquote {\bibinfo {title}
  {Optical binding},}\ }\href {\doibase 10.1103/physrevlett.63.1233} {\bibfield
   {journal} {\bibinfo  {journal} {Physical Review Letters}\ }\textbf {\bibinfo
  {volume} {63}},\ \bibinfo {pages} {1233--1236} (\bibinfo {year}
  {1989})}\BibitemShut {NoStop}%
\bibitem [{\citenamefont {Burns}\ \emph {et~al.}(1990)\citenamefont {Burns},
  \citenamefont {Fournier},\ and\ \citenamefont {Golovchenko}}]{Burns1990}%
  \BibitemOpen
  \bibfield  {author} {\bibinfo {author} {\bibfnamefont {M.~M.}\ \bibnamefont
  {Burns}}, \bibinfo {author} {\bibfnamefont {J.-M.}\ \bibnamefont {Fournier}},
  \ and\ \bibinfo {author} {\bibfnamefont {J.~A.}\ \bibnamefont
  {Golovchenko}},\ }\bibfield  {title} {\enquote {\bibinfo {title} {Optical
  matter: Crystallization and binding in intense optical fields},}\ }\href
  {\doibase 10.1126/science.249.4970.749} {\bibfield  {journal} {\bibinfo
  {journal} {Science}\ }\textbf {\bibinfo {volume} {249}},\ \bibinfo {pages}
  {749--754} (\bibinfo {year} {1990})}\BibitemShut {NoStop}%
\bibitem [{\citenamefont {Grzegorczyk}\ \emph {et~al.}(2006)\citenamefont
  {Grzegorczyk}, \citenamefont {Kemp},\ and\ \citenamefont
  {Kong}}]{Grzegorczyk2006}%
  \BibitemOpen
  \bibfield  {author} {\bibinfo {author} {\bibfnamefont {Tomasz~M.}\
  \bibnamefont {Grzegorczyk}}, \bibinfo {author} {\bibfnamefont {Brandon~A.}\
  \bibnamefont {Kemp}}, \ and\ \bibinfo {author} {\bibfnamefont {Jin~Au}\
  \bibnamefont {Kong}},\ }\bibfield  {title} {\enquote {\bibinfo {title}
  {Stable optical trapping based on optical binding forces},}\ }\href {\doibase
  10.1103/physrevlett.96.113903} {\bibfield  {journal} {\bibinfo  {journal}
  {Physical Review Letters}\ }\textbf {\bibinfo {volume} {96}} (\bibinfo {year}
  {2006}),\ 10.1103/physrevlett.96.113903}\BibitemShut {NoStop}%
\bibitem [{\citenamefont {Metzger}\ \emph {et~al.}(2006)\citenamefont
  {Metzger}, \citenamefont {Dholakia},\ and\ \citenamefont
  {Wright}}]{Metzger2006}%
  \BibitemOpen
  \bibfield  {author} {\bibinfo {author} {\bibfnamefont {N.~K.}\ \bibnamefont
  {Metzger}}, \bibinfo {author} {\bibfnamefont {K.}~\bibnamefont {Dholakia}}, \
  and\ \bibinfo {author} {\bibfnamefont {E.~M.}\ \bibnamefont {Wright}},\
  }\bibfield  {title} {\enquote {\bibinfo {title} {Observation of bistability
  and hysteresis in optical binding of two dielectric spheres},}\ }\href
  {\doibase 10.1103/physrevlett.96.068102} {\bibfield  {journal} {\bibinfo
  {journal} {Physical Review Letters}\ }\textbf {\bibinfo {volume} {96}}
  (\bibinfo {year} {2006}),\ 10.1103/physrevlett.96.068102}\BibitemShut
  {NoStop}%
\bibitem [{\citenamefont {Kar{\'{a}}sek}\ \emph {et~al.}(2006)\citenamefont
  {Kar{\'{a}}sek}, \citenamefont {Dholakia},\ and\ \citenamefont
  {Zem{\'{a}}nek}}]{Karsek2006}%
  \BibitemOpen
  \bibfield  {author} {\bibinfo {author} {\bibfnamefont {V.}~\bibnamefont
  {Kar{\'{a}}sek}}, \bibinfo {author} {\bibfnamefont {K.}~\bibnamefont
  {Dholakia}}, \ and\ \bibinfo {author} {\bibfnamefont {P.}~\bibnamefont
  {Zem{\'{a}}nek}},\ }\bibfield  {title} {\enquote {\bibinfo {title} {Analysis
  of optical binding in one dimension},}\ }\href {\doibase
  10.1007/s00340-006-2297-8} {\bibfield  {journal} {\bibinfo  {journal}
  {Applied Physics B}\ }\textbf {\bibinfo {volume} {84}},\ \bibinfo {pages}
  {149--156} (\bibinfo {year} {2006})}\BibitemShut {NoStop}%
\bibitem [{\citenamefont {Metzger}\ \emph {et~al.}(2007)\citenamefont
  {Metzger}, \citenamefont {Marchington}, \citenamefont {Mazilu}, \citenamefont
  {Smith}, \citenamefont {Dholakia},\ and\ \citenamefont
  {Wright}}]{Metzger2007}%
  \BibitemOpen
  \bibfield  {author} {\bibinfo {author} {\bibfnamefont {N.~K.}\ \bibnamefont
  {Metzger}}, \bibinfo {author} {\bibfnamefont {R.~F.}\ \bibnamefont
  {Marchington}}, \bibinfo {author} {\bibfnamefont {M.}~\bibnamefont {Mazilu}},
  \bibinfo {author} {\bibfnamefont {R.~L.}\ \bibnamefont {Smith}}, \bibinfo
  {author} {\bibfnamefont {K.}~\bibnamefont {Dholakia}}, \ and\ \bibinfo
  {author} {\bibfnamefont {E.~M.}\ \bibnamefont {Wright}},\ }\bibfield  {title}
  {\enquote {\bibinfo {title} {Measurement of the restoring forces acting on
  two optically bound particles from normal mode correlations},}\ }\href
  {\doibase 10.1103/physrevlett.98.068102} {\bibfield  {journal} {\bibinfo
  {journal} {Physical Review Letters}\ }\textbf {\bibinfo {volume} {98}}
  (\bibinfo {year} {2007}),\ 10.1103/physrevlett.98.068102}\BibitemShut
  {NoStop}%
\bibitem [{\citenamefont {Dholakia}\ and\ \citenamefont
  {Zem{\'{a}}nek}(2010)}]{Dholakia2010}%
  \BibitemOpen
  \bibfield  {author} {\bibinfo {author} {\bibfnamefont {Kishan}\ \bibnamefont
  {Dholakia}}\ and\ \bibinfo {author} {\bibfnamefont {Pavel}\ \bibnamefont
  {Zem{\'{a}}nek}},\ }\bibfield  {title} {\enquote {\bibinfo {title}
  {Colloquium: Gripped by light: Optical binding},}\ }\href {\doibase
  10.1103/revmodphys.82.1767} {\bibfield  {journal} {\bibinfo  {journal}
  {Reviews of Modern Physics}\ }\textbf {\bibinfo {volume} {82}},\ \bibinfo
  {pages} {1767--1791} (\bibinfo {year} {2010})}\BibitemShut {NoStop}%
\bibitem [{\citenamefont {Grzegorczyk}\ \emph {et~al.}(2014)\citenamefont
  {Grzegorczyk}, \citenamefont {Rohner},\ and\ \citenamefont
  {Fournier}}]{Grzegorczyk2014}%
  \BibitemOpen
  \bibfield  {author} {\bibinfo {author} {\bibfnamefont {Tomasz~M.}\
  \bibnamefont {Grzegorczyk}}, \bibinfo {author} {\bibfnamefont {Johann}\
  \bibnamefont {Rohner}}, \ and\ \bibinfo {author} {\bibfnamefont {Jean-Marc}\
  \bibnamefont {Fournier}},\ }\bibfield  {title} {\enquote {\bibinfo {title}
  {Optical mirror from laser-trapped mesoscopic particles},}\ }\href {\doibase
  10.1103/physrevlett.112.023902} {\bibfield  {journal} {\bibinfo  {journal}
  {Physical Review Letters}\ }\textbf {\bibinfo {volume} {112}} (\bibinfo
  {year} {2014}),\ 10.1103/physrevlett.112.023902}\BibitemShut {NoStop}%
\bibitem [{\citenamefont {Ng}\ \emph {et~al.}(2005)\citenamefont {Ng},
  \citenamefont {Lin}, \citenamefont {Chan},\ and\ \citenamefont
  {Sheng}}]{Ng2005}%
  \BibitemOpen
  \bibfield  {author} {\bibinfo {author} {\bibfnamefont {Jack}\ \bibnamefont
  {Ng}}, \bibinfo {author} {\bibfnamefont {Z.~F.}\ \bibnamefont {Lin}},
  \bibinfo {author} {\bibfnamefont {C.~T.}\ \bibnamefont {Chan}}, \ and\
  \bibinfo {author} {\bibfnamefont {Ping}\ \bibnamefont {Sheng}},\ }\bibfield
  {title} {\enquote {\bibinfo {title} {Photonic clusters formed by dielectric
  microspheres: Numerical simulations},}\ }\href {\doibase
  10.1103/physrevb.72.085130} {\bibfield  {journal} {\bibinfo  {journal}
  {Physical Review B}\ }\textbf {\bibinfo {volume} {72}} (\bibinfo {year}
  {2005}),\ 10.1103/physrevb.72.085130}\BibitemShut {NoStop}%
\bibitem [{\citenamefont {Lehmberg}(1970)}]{Lehmberg1970}%
  \BibitemOpen
  \bibfield  {author} {\bibinfo {author} {\bibfnamefont {R.~H.}\ \bibnamefont
  {Lehmberg}},\ }\bibfield  {title} {\enquote {\bibinfo {title} {Radiation from
  an {$N$}-atom system. {I}. {G}eneral formalism},}\ }\href {\doibase
  10.1103/PhysRevA.2.883} {\bibfield  {journal} {\bibinfo  {journal} {Phys.
  Rev. A}\ }\textbf {\bibinfo {volume} {2}},\ \bibinfo {pages} {883--888}
  (\bibinfo {year} {1970})}\BibitemShut {NoStop}%
\bibitem [{\citenamefont {Courteille}\ \emph {et~al.}(2010)\citenamefont
  {Courteille}, \citenamefont {Bux}, \citenamefont {Lucioni}, \citenamefont
  {Lauber}, \citenamefont {Bienaim\'e}, \citenamefont {Kaiser},\ and\
  \citenamefont {Piovella}}]{Courteille2010}%
  \BibitemOpen
  \bibfield  {author} {\bibinfo {author} {\bibfnamefont {Ph.~W.}\ \bibnamefont
  {Courteille}}, \bibinfo {author} {\bibfnamefont {S.}~\bibnamefont {Bux}},
  \bibinfo {author} {\bibfnamefont {E.}~\bibnamefont {Lucioni}}, \bibinfo
  {author} {\bibfnamefont {K.}~\bibnamefont {Lauber}}, \bibinfo {author}
  {\bibfnamefont {T.}~\bibnamefont {Bienaim\'e}}, \bibinfo {author}
  {\bibfnamefont {R.}~\bibnamefont {Kaiser}}, \ and\ \bibinfo {author}
  {\bibfnamefont {N.}~\bibnamefont {Piovella}},\ }\bibfield  {title} {\enquote
  {\bibinfo {title} {Modification of radiation pressure due to cooperative
  scattering of light},}\ }\href@noop {} {\bibfield  {journal} {\bibinfo
  {journal} {Euro. Phys. J. D}\ }\textbf {\bibinfo {volume} {58}},\ \bibinfo
  {pages} {69} (\bibinfo {year} {2010})}\BibitemShut {NoStop}%
\bibitem [{\citenamefont {Goetschy}\ and\ \citenamefont
  {Skipetrov}(2011)}]{Goetschy2011}%
  \BibitemOpen
  \bibfield  {author} {\bibinfo {author} {\bibfnamefont {A.}~\bibnamefont
  {Goetschy}}\ and\ \bibinfo {author} {\bibfnamefont {S.~E.}\ \bibnamefont
  {Skipetrov}},\ }\bibfield  {title} {\enquote {\bibinfo {title} {Non-hermitian
  euclidean random matrix theory},}\ }\href {\doibase
  10.1103/physreve.84.011150} {\bibfield  {journal} {\bibinfo  {journal}
  {Physical Review E}\ }\textbf {\bibinfo {volume} {84}} (\bibinfo {year}
  {2011}),\ 10.1103/physreve.84.011150}\BibitemShut {NoStop}%
\bibitem [{\citenamefont {Skipetrov}\ and\ \citenamefont
  {Goetschy}(2011)}]{Skipetrov2011}%
  \BibitemOpen
  \bibfield  {author} {\bibinfo {author} {\bibfnamefont {S~E}\ \bibnamefont
  {Skipetrov}}\ and\ \bibinfo {author} {\bibfnamefont {A}~\bibnamefont
  {Goetschy}},\ }\bibfield  {title} {\enquote {\bibinfo {title} {Eigenvalue
  distributions of large euclidean random matrices for waves in random
  media},}\ }\href {\doibase 10.1088/1751-8113/44/6/065102} {\bibfield
  {journal} {\bibinfo  {journal} {Journal of Physics A: Mathematical and
  Theoretical}\ }\textbf {\bibinfo {volume} {44}},\ \bibinfo {pages} {065102}
  (\bibinfo {year} {2011})}\BibitemShut {NoStop}%
\bibitem [{\citenamefont {Skipetrov}\ and\ \citenamefont
  {Sokolov}(2014)}]{Skipetrov2014}%
  \BibitemOpen
  \bibfield  {author} {\bibinfo {author} {\bibfnamefont
  {S.{\hspace{0.167em}}E.}\ \bibnamefont {Skipetrov}}\ and\ \bibinfo {author}
  {\bibfnamefont {I.{\hspace{0.167em}}M.}\ \bibnamefont {Sokolov}},\ }\bibfield
   {title} {\enquote {\bibinfo {title} {Absence of anderson localization of
  light in a random ensemble of point scatterers},}\ }\href {\doibase
  10.1103/physrevlett.112.023905} {\bibfield  {journal} {\bibinfo  {journal}
  {Physical Review Letters}\ }\textbf {\bibinfo {volume} {112}} (\bibinfo
  {year} {2014}),\ 10.1103/physrevlett.112.023905}\BibitemShut {NoStop}%
\bibitem [{\citenamefont {Skipetrov}(2016)}]{Skipetrov2016}%
  \BibitemOpen
  \bibfield  {author} {\bibinfo {author} {\bibfnamefont {S.~E.}\ \bibnamefont
  {Skipetrov}},\ }\bibfield  {title} {\enquote {\bibinfo {title} {Finite-size
  scaling analysis of localization transition for scalar waves in a
  three-dimensional ensemble of resonant point scatterers},}\ }\href {\doibase
  10.1103/physrevb.94.064202} {\bibfield  {journal} {\bibinfo  {journal}
  {Physical Review B}\ }\textbf {\bibinfo {volume} {94}} (\bibinfo {year}
  {2016}),\ 10.1103/physrevb.94.064202}\BibitemShut {NoStop}%
\bibitem [{\citenamefont {Guerin}\ and\ \citenamefont
  {Kaiser}(2017)}]{Guerin2017}%
  \BibitemOpen
  \bibfield  {author} {\bibinfo {author} {\bibfnamefont {William}\ \bibnamefont
  {Guerin}}\ and\ \bibinfo {author} {\bibfnamefont {Robin}\ \bibnamefont
  {Kaiser}},\ }\bibfield  {title} {\enquote {\bibinfo {title} {Population of
  collective modes in light scattering by many atoms},}\ }\href {\doibase
  10.1103/physreva.95.053865} {\bibfield  {journal} {\bibinfo  {journal}
  {Physical Review A}\ }\textbf {\bibinfo {volume} {95}} (\bibinfo {year}
  {2017}),\ 10.1103/physreva.95.053865}\BibitemShut {NoStop}%
\bibitem [{\citenamefont {Letokhov}\ \emph {et~al.}(1977)\citenamefont
  {Letokhov}, \citenamefont {Minogin},\ and\ \citenamefont
  {Pavlik}}]{Letokhov1976}%
  \BibitemOpen
  \bibfield  {author} {\bibinfo {author} {\bibfnamefont {V.S.}\ \bibnamefont
  {Letokhov}}, \bibinfo {author} {\bibfnamefont {V.G.}\ \bibnamefont
  {Minogin}}, \ and\ \bibinfo {author} {\bibfnamefont {B.D.}\ \bibnamefont
  {Pavlik}},\ }\bibfield  {title} {\enquote {\bibinfo {title} {Cooling and
  capture of atoms and molecules by a resonant light field},}\ }\href
  {http://www.jetp.ac.ru/cgi-bin/e/index/e/45/4/p698?a=list} {\bibfield
  {journal} {\bibinfo  {journal} {Sov. Phys. JETP}\ }\textbf {\bibinfo {volume}
  {45}},\ \bibinfo {pages} {698} (\bibinfo {year} {1977})}\BibitemShut
  {NoStop}%
\bibitem [{\citenamefont {Zhu}\ \emph {et~al.}(2016)\citenamefont {Zhu},
  \citenamefont {Cooper}, \citenamefont {Ye},\ and\ \citenamefont
  {Rey}}]{Zhu2016}%
  \BibitemOpen
  \bibfield  {author} {\bibinfo {author} {\bibfnamefont {Bihui}\ \bibnamefont
  {Zhu}}, \bibinfo {author} {\bibfnamefont {John}\ \bibnamefont {Cooper}},
  \bibinfo {author} {\bibfnamefont {Jun}\ \bibnamefont {Ye}}, \ and\ \bibinfo
  {author} {\bibfnamefont {Ana~Maria}\ \bibnamefont {Rey}},\ }\bibfield
  {title} {\enquote {\bibinfo {title} {Light scattering from dense cold atomic
  media},}\ }\href {\doibase 10.1103/PhysRevA.94.023612} {\bibfield  {journal}
  {\bibinfo  {journal} {Phys. Rev. A}\ }\textbf {\bibinfo {volume} {94}},\
  \bibinfo {pages} {023612} (\bibinfo {year} {2016})}\BibitemShut {NoStop}%
\end{thebibliography}%

\end{document}